# Exploration of Galactic γ-Ray Supernova Remnants

TIAN Wenwu[*]，ZHANG Jianli
National Astronomical Observatories, CAS, Beijing, 100012

New generational very-high-energy telescope arrays have been detecting more than 120 TeV γ-ray sources. Multi-wavelength observations on these Gamma-ray sources have proven to be robust in shedding light on their nature. The coming radio telescope arrays like ASKAP and FAST may find more faint (extended) radio sources due to their better sensitivities and resolutions, might identify more previously un-identified γ-ray sources and set many new targets for future deep surveys by very-high-energy ground-based telescopes like LHAASO. We in the paper summarize a list of known Galactic γ-ray Supernova Remnants (SNRs) with or without radio emissions so far, which includes some SNRs deserving top priority for future multi-wavelength observations.

## 1. Background

Astrophysical shock waves have been considered as efficient accelerators of Cosmic Rays (CRs, mainly composed of nuclei, protons and electrons) since the 1980's [1]. Based on the inferred Galactic CR energy density (1−2 eV/ $cm^3$) and the inferred time (~$6x10^6$ yrs) that an average cosmic ray stays in the Galaxy, the Galactic CRs have a total power of ~$10^{41}$ erg /s. By comparison with the supernova energy and rates (their average explosion energy of ~ $10^{51}$ erg and an inferred Galactic supernova rate of ~ 2 - 3 per century), Galactic Supernova Remnant (SNR) shocks accelerating charged particles in the interstellar medium (ISM) and circum-stellar medium (CSM) are able to produce most of the Galactic cosmic rays up to $10^{15}$ eV . As a key issue to explore the origin of the CSs, we need first understand the relative efficiency of acceleration of protons versus electrons as well as the maximum energies obtained.

CRs are mainly composed of charged particles which are deflected in the magnetic fields, so direct observations on the charged particles carry no information on their origin. However, observations of Gamma-ray photons as small components of the CRs (<1%) play a key role in this issue due to two reasons: Gamma-rays have straight-line propagation; Gamma-rays may be produced by relativistic particles' interactions.

New generational very-high-energy telescope arrays have detected more than 120 TeV γ-ray sources in the Universe in recent years. Seeking low-energy band counterparts of TeV sources plays a key role in shedding light on the nature of the TeV sources. Two major γ-ray generators have been identified now, i.e. young shell-type SNRs and pulsar wind nebulae [2,3]. There are also other possible candidates as counterparts to some unidentified TeV sources[4, 5,6].

[*] Corresponding author (email:tww@bao.ac.cn)

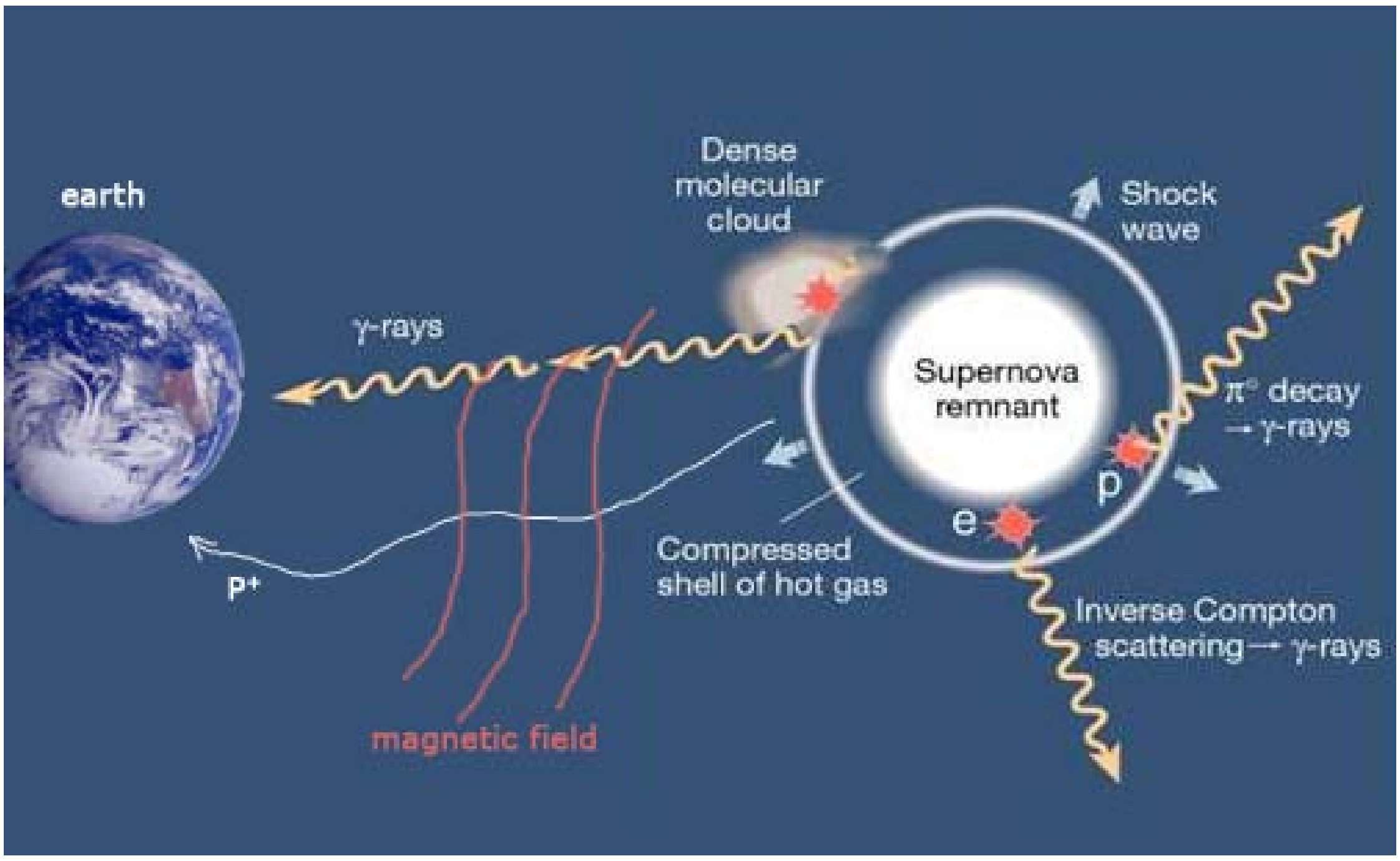

Fig. 1 diagrams that γ-rays from cosmic ray sources are straight-line transportation while charged particles are deflected in ISM magnetic fields on their travel from the origin to the Earth. Two main γ-ray emission mechanisms here are: hadronic origin and leptonic origin (for details please refer to the text).Credits go to  Mr. SU Hongquan for helping produce Figure 1 which in turn benefits from Dr. Aharonian's paper (http://www.nature.com/nature/journal/v416/n6883/full/416797a.html).

With the advent of new general X-ray and γ-ray observatories (e.g., XMM-NEWTON, Chandra, HESS, Fermi etc), recent observations have shown that SNR shocks are able to accelerate particles to TeV energies[7,2,8]. Theoretical exploration has revealed that two primary TeV photon emission mechanisms play major roles in the processes, i.e., the decay of neutral pions ($\pi^0$) by p-p interactions and the inverse-Compton scattering of electrons on ambient photons. These particles may or may not immediately escape the acceleration and emission sites due to the presence of magnetic fields in SNR, which depends on the energy threshold of particles. Part of the charged particles may be detected after they escape to the interstellar space from the accelerating sites when the MHD disturbance which constrains the particles gets decreased. However, it is still extremely difficult for us to trace their accelerating sites by detecting them because Galactic magnetic fields have changed their directions of transmission. However, the TeV photons tell us different story because they escape directly and transmit in a straight-line way, which makes it possible to find the sites of these processes.

Observations in the radio, X-ray and γ-ray wavebands are so far the best way to access non-thermal acceleration processes in SNR. Although studying γ-rays emissions may point to the sites of the particle acceleration, the particles are either hadrons or leptons due to existing two main mechanisms. So far, we might have detected the two classes of TeV γ-ray photons from SNR, but we cannot claim that we have solved the CRs' origin issue before we may clearly distinguish them.

Studying evolved SNRs of older than $10^3$ yr seems bring new hopes because evolved SNRs have very different characteristics comparing to young SNRs. [4] have revealed that the γ- and X-ray emissions in an evolved SNR are expected to be dominated by the decay of neutral pions and by the synchrotron radiation of secondary electrons from charged pion decay, respectively.. So detecting both γ- and X-ray emissions from many faint radio SNRs might provide key evidence that Galactic CRs originates from SNRs' accelerating charged particles.

Recently, we have given evidence of CRs originating from old SNR using multi-wave observations. Based on a new distance-measurement method, [9] provided observational evidence supporting that the old SNR W41 encounters a GMC to emit TeV γ-rays (HESS J1834-087). In this case, the CR protons could diffuse only partway into the cloud, resulting in p-p to π0 decay TeV γ-rays, so TeV γ-rays in W41 primarily have hadronic origin. Further, [10] discovered a new faint SNR G353.6-0.7 with age of 27000 yrs, whose radio and X-ray morphologies closely match the outline of HESS J1731-347, one previously unidentified extended TeV source. Deep research using new results from XMM-Newton and SUZAKU X-ray observations, and Delinha CO observations demonstrated that G353.6-0.7 may be coincident with CO clouds. Based on the probable association between the X-ray and gamma-ray emissions and likely association between the CO cloud and the SNR, [8] argued that the extended TeV emission originates from the interaction between the SNR shock and the adjacent CO clouds.

Our study has shown that it is highly likely that evolved SNRs, as well as young SNRs, play a key role in proving that SNRs are the source of the Galactic CRs. Especially, almost half of the detected Galactic TeV γ -rays sources so far have no low energy-band's counterparts. For upcoming very-high-energy telescope surveys, more such unidentified TeV sources will be found （e.g. CTA, LHAASO）. To identify the unidentified TeV sources becomes increasingly important. However, it is difficult to do this by using current radio and X-ray instruments as a result of their limited sensitivity, if the TeV sources are faint SNRs. In fact, there is a discrepancy between the number of known SNRs (276) and the number predicted by theory (above 2000 Galactic SNRs). This has been considered the result of selection effects in current sensitivity-limited surveys, which favor the discovery of the brighter remnants (e.g. [11]). Survey data from observations that combine high sensitivity and resolution with low radio frequencies are becoming the rich modern hunting-grounds for new SNRs. Some new SNRs have been discovered from the 90 cm Very Large Array survey of the inner Galaxy [12]，the Canadian Galactic Plane Survey (CGPS) of the outer Galaxy[13,14]，the Southern Galactic Plane Survey [10], and a Sino-German 6 cm polarization survey of the Galactic plane[15] .

## 2. Current Statistics Results

New general space and ground-based high-energy telescopes have detected many gamma-ray sources. Most Galactic gamma-ray sources are probably associated with SNRs/PWNs. Table 1 shows all Galactic TeV SNRs/PWNs (Table 1) and Table 2 is the Galactic GeV SNRs/PWNs and unidentified TeV sources, based on all accessible observations and references so far. We focus on the GeV/TeV sources with radio emissions. We have only selected the SNRs with GeV/TeV emission in the catalog compiled by Ferrand & Safi-Harb[16] (see http://www.physics.Umanitoba.ca/snr/SNRcat), which includes all the high energy SNRs in the Galaxy. We have added the TeV sources not appearing in Ferrand & Safi-Harb's catalog but appearing in the TeV catalog (see http://tevcat.uchicago.edu). Some TeV sources have not been identified by other wave band observations so far. We have also given/calculated fluxes of the Gev/TeV SNRs with the unit of Crab. In addition, we have made notes if a SNR is in the LHAASO's field of view as a bonus for the LHAASO design. In the tables, the second column is the "SNR ID Gal. name" which uniquely identifies the object (with its Galactic coordinates). The columns "RA(2000)" and "Dec(2000)" give the location (Right Ascension and Declination of the J2000). The column "LHAASO" shows whether the SNR/PWN is in the field of view of LHAASO. The sixth column "name (Gamma-ray)" are gamma-ray source names. The seventh column is the morphology type. Columns eight and nine are the age and distance. column twelve "SNR/MC" shows if there exists an interaction between SNR shock with adjacent molecular clouds. Symbol "?" hints an uncertainty in TeV sources' classification and association relationship.

2.1. **Current Catalogue of Gamma-Ray/SNR Associations**: **GeV/TeV Galactic SNR**

Table 1 lists 61 TeV gamma ray sources, which are likely identified as SNRs/PWNe. Most of them are discovered by H.E.S.S., 21 cases of which are in LHAASO field of view. 28 cases of TeV sources have likely shell structures (20 are considered certain, and 8 have not been identified). 32 of TeV sources have been considered as PWN (28 are certain, and 4 are possible). 28 cases are composite (11 are confirmed, and 17 are possible). Association with a molecular cloud has been reported in 28 cases, and 4 cases are considered unlikely. Most of them with measured distances are nearby, i.e. 32 of 52 cases have distances less than 6 kpc. Many SNRs are very young, i.e. 17 of 39 cases with known ages have ages less than 3 k years.

**Table 1: TeV Galactic SNRs/PWNe which have radio or X-ray counterparts**

| | SNR ID Gal. name | RA (2000) | Dec (2000) | LHAASO | Name (Gamma-ray) | Type | Age (kyr) | Distance (kpc) | Radio size | Gamma-Flux (Crab Units) | SNR/ MC | Refs |
|---|---|---|---|---|---|---|---|---|---|---|---|---|
| 1 | G0.0+0.0(SGR A East) | 17h 45m 44s | -29º 00' 00" | N | HESS J1745-290(GT) | S? | 8 | 8.5 | 2.5'-3.5' | 0.05(≥0.165TeV)(H,V,C,MA,F) | Y | 17, 18, 19, 20, 21, 22, 23, 24, 25, 26, 27 |
| 2 | G0.9+0.1 | 17h 47m 21s | -28º 09' 00" | N | HESS J1747-281(T) | C PWN | 1.9(S), 5(P) | 10(S),13(P) | 8' | 0.02(≥0.2TeV)(H) | | 17, 28 |
| 3 | G5.7-0.1? | 17h 58m 49s | -24º 03' 00" | N | HESS J1800-240C(T) | | 35–45 | | 9' – 12' | 0.01(0.3-5TeV)(H) | Y | 17, 29 |
| 4 | G5.9-0.37 | 18h 00m 26.4s | -24º 02' 20.4" | N | HESS J1800-240B(GT) | S? C? | 35–45 | | | 0.02(0.3-5TeV)(H,F) | | 17, 29 |
| 5 | G6.14-0.63 | 18h 01m 57.8s | -23º 57' 43.2" | N | HESS J1800-240A(GT) | S? C? | 35–45 | | | 0.02(0.3-5TeV)(H,F) | | 17, 29 |
| 6 | G6.4-0.1( W28) | 18h 00m 30s | -23º 26' 00" | N | (GT) | C | 35–45(S) | 1.8 - 1.9(S) | 48' | (H) | Y | 17, 29 |
| 7 | G6.5-0.4 | 18h 02m 11s | -23º 34' 00" | N | HESS J1801-233(T) | S | | | 18' | 0.02(0.3-5TeV)(H) | | 17, 29 |
| 8 | G8.7-0.1 (W30) | 18h 05m 30s | -21º 26' 00" | N | HESS J1804-216(GT) | S? PWN | 10–50 | 4(S,P) | 45' | 0.25(≥0.2TeV)(H,C,F) | Y | 17, 30 |
| 9 | G11.2-0.3 | 18h 11m 27s | -19º 25' 00" | N | HESS J1809-193(T) | C | ≤2(S), 24(P) | ≥ 5(S) 5(P) | 4' | 0.14(≥0.25TeV)(H) | N | |
| 10 | G12.8-0.0(W33) | 18h 13m 37s | -17º 49' 00" | N | HESS J1813-178(T) | C? PWN | 1.2(S),5(P) | 4.7(P) | 3' | 0.06(≥0.2TeV)(H,MA) | N | |
| 11 | G15.4+0.1 | 18h 18m 02s | -15º 27' 00" | N | HESS J1818-154(T) | S? | | | 14'-15' | 0.018(≥1TeV)(H) | | 31 |
| 12 | G18.0-0.7 | 18h 25m 41s | -13º 50' 20" | N | HESS J1825-137(GT) | PWN | | 4(P) | | 0.17(≥0.2TeV)(H,F) | | 32, 33, 34 |
| 13 | G21.5-0.9 | 18h 33m 33s | -10º 35' 00" | N | HESS J1833-105(T) | C PWN | 0.7-1.07(S),4.8(P) | 4.7(S,P) | 4' | 0.02(≥0.2TeV)(H) | Y? | 17, 35 |
| 14 | G22.7-0.2 | 18h 32m 46.83s | -09º 21' 54.5" | Y | HESS J1832-093(T) | S? | | | 26' | 0.007(≥1TeV)(H) | Y? | 36 |
| 15 | G23.3-0.3(W41) | 18h 34m 45s | -08º 48' 00" | Y | HESS J1834-087(GT) | S? PWN? | 100(S) | 3.9-4.5(S) | 27' | 0.08(≥0.2TeV)(H,MA,F) | Y? | 17, 37, 38, |
| 16 | G25.5+0.0 | 18h 37h 38.4h | -06º 57' 00" | Y | HESS J1837-069(T) | PWN? | | | | 0.132(≥0.2TeV)(H) | | 32 |
| 17 | G29.7-0.3(Kes 75) | 18h 46m 25s | -02º 59' 00" | Y | HESS J1846-029(T) | C PWN | 0.9-4.3(S), 0.725 (P) | 5.1-7.5(Tian) | 3' | 0.02(≥0.25TeV)(H) | Y | 17, 39 |
| 18 | G32.8-0.1(Kes 78) | 18h 51m 25s | -00º 08' 00" | Y | HESS J1852-000(T) | S? | | 4.8(S) | 17' | ?(H) | Y | 17, 40, 41 |
| 19 | G35.6-0.4 ? | 18h 54m 53s | 02º 33' 25" | Y | HESS J1858+020(T) | S | 30(S),160(P) | 3.7-10.5(S),8(P) | 11'-15' | 0.018(0.5-80TeV)(H) | | 42, 43 |
| 20 | G40.5-0.5 | 19h 07m 10s | 06º 31' 00" | Y | HESS J1908+063,MGRO 1908+06 (GT) | S | 20(P) | 3.2(P) | 22' | 0.17(≥1TeV)(H,V,MI,A,F) | Y? | 44, 45 |
| 21 | G43.3-0.2(W49B) | 19h 11m 08s | 09º 06' 00" | Y | HESS J1911+090(GT) | C | 1-4(S) | 8 – 11(S) | 3' - 4' | 0.005(≥0.26TeV)(H,F) | Y | 37, 46, 47 |
| 22 | G49.2-0.7 (W51) | 19h 23m 50s | 14º 06' 00" | Y | HESS J1923+141(GT) | C? PWN | 10(S) | 5.5 – 6(S) | 30' | 0.003(≥1TeV)(H,MA,F) | Y | 17, 48, 49, 50, 51 |
| 23 | G54.1+0.3 | 19h 30m 31s | 18º 52' 00'' | Y | (T) | PWN | 2.5-3.3(S), 2.9(P) | 5.6-7.2(S),5(P) | 1.5' | 0.025(≥1TeV)(V) | Y? | |
| 24 | G65.1+0.6 | 19h 54m 40s | 28º 35' 00'' | Y | 0FGL J1954.4+2838(GT) | S | 4-14 | 9.2(S) | 50'-90' | 0.23(=35TeV)(MI,F) | | |
| 25 | G74.9+1.2(CTB 87) | 20h 16m 02s | 37º 12' 00'' | Y | VER J2016+372(GT) | C? PWN | | 6.1-12(S) | 6' - 8' | 0.01(≥1TeV)(V,A) | Y? | 52, 53, 54 |
| 26 | G75.2+0.1(Cisne) | 20h18m 35.03s | 36º50'00'' | Y | MGRO J2019+37(T) | PWN ? | | ≥10(S),4(P) | | 0.67(=35TeV)(MI) | | 52, 55, 53, 56 |
| 27 | G78.2+2.1(gammaCygni) | 20h 20m 50s | 40º 26' 00'' | Y | VER J2019+407(GT) | S | 6.6(S) | 1.5(S) | 60' | ?(V,F) | Y? | 17, 57, 58, 59, 60 |
| 28 | G106.3+2.7(Boomerang) | 22h 27m 30s | 60º 50' 00'' | Y | 2FGL J2229.0+6114(GT) | C? PWN | 10(P) | 0.8(S) ,3(P) | 24' - 60' | 0.05 (≥1TeV)(V,MI,F) | Y? | 17, 59, 61 |
| 29 | G111.7-2.1( Cas A) | 23h 23m 26s | 58º 48' 00'' | Y | 2FGL J2323.4+5849(GT) | S | 0.316-0.352(S) | 3.3 - 3.7(S) | 5' | 0.03(≥1TeV)(V,MA,F) | Y? | 17, 62 |
| 30 | G119.5+10.2 | 00h 06m 40s | 72º 45' 00'' | N | CTA 1(GT) | S PWN | 13-17 | 1.4(S) | 90' | 0.04(≥1TeV)(V,F) | | 17, 63, 1, 64 |
| 31 | G120.1+1.4 | 00h 25m 18s | 64º 09' 00'' | Y | Tycho(GT) | S | 0.440 | 2.5-5(S) | 8' | 0.009 (MA,F) | Y? | 57, 58, 65, 66, 67, 68, 69, 70 |
| 32 | G184.6-5.8 | 05h 34m 31s | 22º 01' 00'' | Y | Crab(GT) | C? PWN | 0.958 | 1.5-2.5(S), 2(P) | 5' - 7' | 1(H,V,MA,MI,A,F) | N | 17, 71, 72, 73 |
| 33 | G189.1+3.0 (IC443) | 06h 17m 00s | 22º 34' 00'' | Y | MAGIC J0616+225(GT) | C PWN | 30(S,P) | 0.7 – 2(S) | 45' | 0.03 (V,MA,MI,F) | Y | 17, 74 |
| 34 | G195.1+4.3 | 06h 32m 28s | 17º 22' 00'' | Y | Geminga (GT) | C? PWN | | 0.25(P) | | 0.23(35TeV)(MI,F) | | 35 |
| 35 | G205.5 +0.5(Monoceros) | 06h 39m 00s | 06º 30' 00" | Y | HESS J0632+057(GT) | S | 30-150 | 0.8 - 1.6(S) | 220' | 0.03(≥0.4TeV)(H,MA,V,F) | Y? | 17 |
| 36 | G263.9-3.3(Vela X) | 08h 34m 00s | -45º 50' 00" | N | HESS J0835-455(GT) | C PWN | | 0.25-0.3(S), 0.29(P) | 255' | 0.75(≥0.45TeV)(H,F) | Y? | 17, 35, 75 |
| 37 | G266.2-1.2(Vela Jr.,RX J0852.0-4622) | 08h 52m 00s | -46º 20' 00" | N | HESS J0852-463(GT) | S | 0.6-4(S) | 0.2 - 2.4(S) | 120' | 1(0.3-20TeV)(H,C,F) | ? | 17, 76, 77, 56 |
| 38 | G284.3-1.8(MSH 10-53) | 10h 18m 15s | -59º 00' 00" | N | HESS J1018-589(GT) | S PWN | 10(S) | 2.9(S),3(P) | 24' | ?(H,F) | Y | 17, 78 |
| 39 | G292.2-0.5 | 11h 19m 20s | -61º 28' 00" | N | HESS J1119-614(T) | S PWN | ≤ 1.7(S) 1.6(P) | 8-8.8(S), 1.6(P) | 15'-20' | 0.04(0.5-10TeV)(H) | | 17 |
| 40 | G304.1-0.2 | 13h 03m 0.4s | -63 11 55 | N | HESS J1303-631(T) | C? PWN | | 7(P) | | 0.17(≥0.38TeV)(H) | | 79, 80 |
| 41 | G309.9-2.5 | 13h 56m 00s | -64º 30' 00" | N | HESS J1356-645(GT) | C? PWN | | 2.5(P) | | 0.11(1-10TeV)(H,F) | | 81, 82, 83 |
| 42 | G313.3 +0.1(Rabbit) | 14h 18m 04s | -60º 58' 31" | N | HESS J1418-609(GT) | C? PWN | | 5.6 | | 0.06(≥0.3TeV)(H,F) | | 44 |
| 43 | G313.6 +0.3(Kookaburra) | 14h 20m 09s | -60º 45' 36" | N | HESS J1420-607(GT) | C? PWN | 13(P) | 5.6(P) | | 0.07(≥0.3TeV)(H,F) | | 44 |
| 44 | G315.4-2.3(RCW 86) | 14h 43m 00s | -62º 30' 00" | N | HESS J1442-624(T) | S | 2-10(S) | 2.4 - 3.2(S) | 42' | 0.1(≥0.48TeV)(H) | ? | 17, 84, 85 |
| 45 | G318.2 +0.1 | 14h 54m 50s | -59º 04' 00" | N | HESS J1457-593(T) | S | | 3.5-9.2(S) | 35'-40' | ?(H) | | 17, 86 |
| 46 | G320.4-1.2(RCW 89,MSH 15-52) | 15h 14m 30s | -59º 08' 00" | N | HESS J1514-591(GT) | C PWN | 1.9(S),1.7(P) | 3.8-6.6(S), 5(P) | 35' | 0.15(≥0.28TeV)(H,C,F) | | 17, 32, 44 |
| 47 | G327.1-1.1 | 15h 54m 25s | -55º 09' 00" | N | (T) | C | 11(S) | 9(S) | 18' | 0.015(1-10TeV)(H) | | 17, 87 |
| 48 | G327.6+14.6(SN 1006) | 15h 02m 50s | -41º 56' 00" | N | HESS J1502-421/HESS J1504-418(T) | S | 1.006 | 2.1 - 2.3(S) | 30' | 0.01 (≥1TeV)(H) | N | 17 |
| 49 | G332.5-0.3 | 16h 16m 24s | -50º 53' 60" | N | HESS J1616-508(T) | C? PWN | | 6.5(P) | | 0.19(≥0.2TeV)(H) | | 79, 80 |
| 50 | G335.2+0.1 | 16h 27m 45s | -48º 47' 00" | N | HESS J1626-490(GT) | S | | | 21' | 0.16(0.6-50TeV)(H) | | 17, 88 |
| 51 | G337.2+0.1 | 16h 35m 55s | -47º 20' 00" | N | HESS J1634-472(GT) | C? PWN? | 1.5(S) | 14(S) | 2' - 3' | 0.12(≥0.2TeV)(H) | | 17 |
| 52 | G338.3-0.0 | 16h 41m 00s | -46º 34' 00" | N | HESS J1640-465(GT) | C? PWN | 4.5-8.2(S) | 8 – 13(S) | 8' | 0.09(≥0.2TeV)(H,F) | | 17 |
| 53 | G343.1-2.3 | 17h 08m 00s | -44º 16' 00" | N | HESS J1708-443(T) | C? PWN | 17.5(P) | 2(P) | 32' | 0.17(1-10TeV)(H,C) | | 17, 89 |
| 54 | G344.7-0.1 | 17h 03m 51s | -41º 42' 00" | N | HESS J1702-420(T) | C? | | 14(S),5(P) | 10' | 0.07(≥0.2)(H) | Y? | 17, 90, 91, 92 |
| 55 | G345.7-0.2 | 17h 07m 20s | -40º 53' 00" | N | HESS J1708-410(T) | S | | | 6' | 0.04(≥0.2TeV)(H) | | 17, 79, 80 |
| 56 | G347.3-0.5 | 17h 13m 50s | -39º 45' 00" | N | RX J1713.7-3946(GT) | S? | 1-10(S) | 1 – 6(S) | 55'-65' | 0.66(≥1TeV)(H,C,F) | Y | 17, 56, 93, 94, 95, 96 |
| 57 | G348.5+0.1(CTB 37A) | 17h 14m 06s | -38º 32' 00" | N | HESS J1714-385(GT) | S PWN | 1-3(S) | 6.3-9.59(S) | 15' | 0.03(≥1TeV)(H,F) | Y | 17, 97, 98 |
| 58 | G348.7 +0.3(CTB 37B) | 17h 13m 55s | -38º 11' 00" | N | HESS J1713-381(T) | S | 0.35–3.15(S),0.95(P) | 13.2 (S) | 55' - 65' | 0.018(≥0.2TeV)(H) | Y | 17, 97, 99 |
| 59 | G348.9-0.4 | 17h 18m 07s | -38º 33' 00" | N | HESS J1718-385(T) | C? PWN | | 4(P) | | 0.02(1-10TeV)(H) | | |
| 60 | G353.6-0.7 | 17h 32m 00s | -34º 44' 00" | N | HESS J1731-347(T) | S | 27(S) | 2.4 - 4 (S) | 30' | 0.16(1-10TeV)(H) | Y? | 17, 100 |
| 61 | G359.1-0.5 | 17h 45m 30s | -29º 57' 00" | N | HESS J1745-303(T) | C | 20–50 | | 24' | 0.05(≥0.2TeV)(H) | Y | 17, 79, 80, 101, 102 |

LHAASO: “N” means the source is not in the field of view of LHAASO. “Y” means it should be detected by LHAASO.

Name (Gamma-ray)：GT: emissions are both measured by Fermi in GeV and other ground-based telescopes in TeV; G: emission is measured only by Fermi in GeV; T: emission is measured only by ground-based telescope(s) in TeV.

Type: Composite=C, Shell = S, Pulsar Wind Nebula = PWN.

Age/Distance: Shell=S, Pulsar=P. Distances of CTB 37A/B are from Tian &Leahy (2012, MNRAS, online earlier version).

Gamma-Flux (Crab Units): H=HESS, V=VERITAS, MA=MAGIC, C=CANGAROO, MI=Milagro, A=ARGO-YBJ, F=FERMI, HE=HEGRA, W=Whipple. Fermi measurements are in the energy range of 1-100GeV. Crab Units: Crab flux[错误！未定义书签。]: differential: 3.45E-11*(E/1TeV)^(-2.63) cm−2s−1TeV−1 .

SNR/Molecular Cloud: “Y” = Yes, “Y?” =likely, “N” = unlikely.

The data used here are mostly from website catalogs: http://www.physics.umanitoba.ca/snr/SNRcat, http://tevcat.uchicago.edu and Caprioli D. [103]. The SNR/MC interactions listed in column 12 are gotten from Ferrand & Safi-Harb’s catalog, also referencing the paper of Jiang B. et al. [104].

### 2.2 Unidentified GeV-TeV Galactic Sources

There are 61 GeV-TeV gamma ray sources in this list: 19 are unidentified TeV objects, and 42 are identified as SNRs discovered by Fermi Gamma-ray LAT. In 30 cases, the object has shell: 25 are considered certain, 5 have not been confirmed. 8 cases have been considered as PWN: 6 are certain, and 2 are possible. 12 cases are composite: 7 are confirmed, and 5 are possible. Association with a molecular cloud has been reported in 12 cases in GeV cases.

## Table 2: Galactic GeV SNRs/PWN and no X-ray or radio counterpart TeV sources

| | SNR ID Gal. name | RA (2000) | Dec (2000) | LHAASO | Name (Gamma-ray) | Type | Age (kyr) | Distance (kpc) | Radio size | Gamma-Flux (Crab Units) | SNR/MC | Refs |
|---|---|---|---|---|---|---|---|---|---|---|---|---|
| 1 | G5.4-1.2?(Milne 56) | 18h 02m 10s | -24º 54' 00" | N | 2FGL J1802.3-2445c(G) | C? | | 4.3-4.5(S), 5(P) | 35' | 0.024(F) | Y | 17 |
| 2 | G11.4-0.1 | 18h 10m 47s | -19º 05' 00" | N | 2FGL J1811.1-1905c(G) | S? | | | 8' | 0.017(F) | | 17 |
| 3 | G20.0-0.2 | 18h 28m 07s | -11º 35' 00" | N | 2FGL J1828.3-1124c(G) | C PWN? | | | 10' | 0.024(F) | | 17 |
| 4 | G21.85-0.11 | 18h 31m 25s | -09º 54' 00" | Y | HESS J1831-098(T) | PWN | | | | 0.04(≥1TeV)(H) | | 105 |
| 5 | G24.7+0.6 | 18h 34m 10s | -07º 05' 00" | Y | 2FGL J1834.7-0705c(G) | C? PWN? | | | 15'-30' | 0.048(F) | | 17 |
| 6 | G26.8-0.2 | 18h 40m 55s | -05º 33' 00" | Y | HESS J1841-055(T) | | | | | 0.37(0.54-80TeV).(H) | | 79, 80 |
| 7 | G27.4+0.0(Kes 73) | 18h 41m 19s | -04º 56' 00" | Y | 2FGL J1841.2-0459c(G) | S | 1.1-1.5(S), 4.3(P) | 7.5-9.1(S) | 4' | 0.055(F) | | 17 |
| 8 | G27.8+0.6 | 18h 39m 50s | -04º 24' 00" | Y | 2FGL J1840.3-0413c(G) | C | | | 30'-50' | 0.032(F) | | 17 |
| 9 | G28.8+1.5 | 18h 39m 00s | -02º 55' 00" | Y | 2FGL J1839.7-0334c(G) | S? | | | 100' | 0.014(F) | | 17 |
| 10 | G29.3+0.51 | 18h 43m 00s | -03º 00' 00" | Y | HESS J1843-033(T) | | | | | ?(H) | | |
| 11 | G31.9+0.0 | 18h 49m 25s | -00º 55' 00" | Y | 2FGL J1849.3-0055(G) | S | 3.7-4.4(S) | ≥ 7.2(S) | 5' - 7' | 0.043(F) | Y | 17 |
| 12 | G32.4+0.1 | 18h 50m 05s | -00º 25' 00" | Y | 2FGL J1850.7-0014 | S | | | 6' | 0.023(F) | | 17 |
| | | | | | c(G) | | | | | | | |
| 13 | G32.64+0.53(IGR J18490-0000) | 18h49m 01.63s | -00° 01' 17.2" | Y | HESS J1849-000(T) | PWN | | 7 | | 0.015(≥0.35TeV)(H) | | |
| 14 | G33.6+0.1(Kes 79) | 18h 52m 48s | 00º 41' 00" | Y | 2FGL J1852.7+0047c(G) | S | 3-15(S) | 7.1(S) | 10' | 0.01(F) | Y? | 17 |
| 15 | G34.7-0.4(W44, 3C392) | 18h 56m 00s | 01º 22' 00" | Y | 2FGL J1855.9+0121e(G) | C PWN | ≥ 10(S), 20(P) | 3(S,P) | 27'-35' | 0.436(F) | Y | 17, 106 |
| 16 | G35.96-0.06 | 18h 57m 11s | 02° 40' 00" | Y | HESS J1857+026(T) | | | | | 0.21(0.6-80TeV) | | 49, 79, 80, 107 |
| 17 | G44.39-.07 | 19h 12m 49s | 10° 09' 06" | Y | HESS J1912+101(T) | PWN? | | | | 0.09 (1-10TeV)(H) | | |
| 18 | G45.7-0.4 | 19h 16m 25s | 11º 09' 00" | Y | 2FGL J1916.1+1106(G) | S | | | 22' | 0.007(F) | | 17 |
| 19 | G54.4-0.3(HC40) | 19h 33m 20s | 18º 56' 00" | Y | 2FGL J1932.1+1913(G) | S | | | 40' | 0.042(F) | Y | 17 |
| 20 | G59.2-4.7(Black Widow) | 19h 59m 35.8s | 20º 47' 28" | Y | 2FGL J1959.5+2047 (G) | C? PWN | | 2.5(P) | | 0.015(F) | | 17, 108 |
| 21 | G65.85-0.23 | 19h 58m 07.61s | 20° 48' 11.9" | Y | 0FGL J1958.1+2848 | PWN | | | | 0.21(Mi) | | |
| 22 | G74.0-8.5(Cygnus Loop) | 20h 51m 00s | 30º 40' 00" | Y | 2FGL J2051.0+3040e(G) | S | 14(S) | 0.4-0.5(S) | 160'-230' | 0.062(F) | Y? | 17 |
| 23 | G76.9+1.0 | 20h 22m 20s | 38º 43' 00" | Y | 2FGL J2022.8+3843c(G) | C? | 9(P) | 8(S) | 9' | 0.013(F) | | 17, 109 |
| 24 | G79.72+1.26 | 20h 29m 38.4s | 41° 11' 24" | Y | MGRO J2031+41(T) | | | | | 0.39(=35 TeV)(MI,A) | | 45, 52, 55, 110 |
| 25 | G80.25+1.07 | 20h 32m 07s | 41° 30' 30" | Y | TeV J2032+4130(T) | | | | | 0.03(=35 TeV)(HE,W,MI,A) | | 52, 55, 110 |
| 26 | G89.0+4.7(HB21) | 20h 45m 00s | 50º 35' 00" | Y | 2FGL J2041.5+5003, J2043.3+5105, J2046.0+4954(G) | C | 19(S) | 0.8(S) | 90'-120' | 0.007 0.01 0.013 (F) | Y | 17 |
| 27 | G114.3+0.3 | 23h 37m 00s | 61° 55' 00" | Y | 2FGL J2333.3+6237(G) | S | 4.1(S) | 0.7(S) | 55'-90' | 0.006(F) | Y | 17 |
| 28 | G116.5+1.1 | 23h 53m 40s | 63° 15' 00" | Y | 2FGL J2358.9+6325(G) | S | 280(S) | 1.6(S) | 60' - 80' | 0.005(F) | | 17 |
| 29 | G132.7+1.3 (HB3) | 02h 17m 40s | 62° 45' 00" | Y | 2FGL J0214.5+6251c,J0218.7+6208c,J0221.4+6257c(G) | S | | 2.2(S) | 80' | 0.006 0.013 0.016 (F) | Y? | 17 |
| 30 | G140.25-16.75 | 02h 23m 12s | 43° 00' 42" | Y | MAGIC J0223+403(T) | | | | | 0.022(≥0.15TeV)(MA | | 111 |
| 31 | G160.9+2.6 (HB9) | 05h 01m 00s | 46° 40' 00" | Y | 2FGL J0503.2+4643(G) | C | | 1.5-4(S) | 120'-140' | 0.006(F) | Y? | 17 |
| 32 | G166.0+4.3 (VRO 42.05.01) | 05h 26m 30s | 42° 56' 00" | Y | 2FGL J0526.6+4308(G) | S | | 4.5(S) | 35'-55' | 0.004(F) | Y? | 17 |
| 33 | G179.0+2.6 | 05h 53m 40s | 31° 05' 00" | Y | 2FGL J0553.9+3104(G) | S? | | | 70' | 0.014(F) | | 17 |
| 34 | G180.0-1.7(S147) | 05h 39m 00s | 27º 50' 00" | Y | 2FGL J0538.1+2718 (G) | S PWN | 26-34 (S),620(P) | 0.36-0.88 (S),1.47(P) | 180' | 0.007(F) | | 17 |
| 35 | G201.3+0.51 | 06h 31m 49.22s | 10º 34' 12.7" | Y | 0FGL J0631.8+1034 | PWN | | 6.55(P) | | 0.29(Mi) | | |
| 36 | G260.4-3.4(Puppis A) | 08h 22m 10s | -43º 00' 00" | Y | 2FGL J0821.0-4254, J0823.0-4246, J0823.4-4305 | S | 3.4(S) | 2.2 | 50'-60' | 0.008 0.031 0.007 (F) | | 17, 112 |
| 37 | G284.8-0.52 | 10h 26m 38.4s | -58º 12' 00" | N | HESS J1026-582(T) | PWN | | | | 0.032(≥0.8TeV)(H) | | 113 |
| 38 | G287.4+0.6 (Puppy) | 10h 48m 16.7s | -58º 31' 48" | N | 2FGL J1048.2-5831(G) | C? PWN | | 3(P) | | 0.154(F) | | |
| 39 | G291.0-0.1(MSH 11-62) | 11h 11m 54s | -60º 38' 00" | N | 2FGL J1112.1-6040 (G) | C PWN | 1.3-10(S) | 1-11(S) | 13'-15' | 0.069(F) | | 17 |
| 40 | G298.6-0.0 | 12h 13m 41s | -62º 37' 00" | N | 2FGL J1214.0-6237(G) | S | | | 9' - 12' | 0.038(F) | | 17 |
| 41 | G304.6+0.1 (Kes 17) | 13h 05m 59s | -62º 42' 00" | N | (G) | S | | 9.7(S) | 8' | ?(F) | Y | 114 |
| 42 | G314.41-0.14 | 14h 27m 52s | -60º 51' 00" | N | HESS J1427-608(T) | | | | | 0.05(0.97-50TeV)(H) | | 79, 80 |
| 43 | G315.1+2.7 | 14h 24m 30s | -57º 50' 00" | N | 2FGL J1411.9-5744(G) | S | | | 150'-190' | 0.007(F) | | 17 |
| 44 | G316.3-0.0(MSH 14-57) | 14h 41m 30s | -60º 00' 00" | N | 2FGL J1441.6-5956(G) | S | | ≥ 7.2(S) | 14'-29' | 0.02(F) | | 17 |
| 45 | G317.95-3.49 | 15h 06m 52.8s | -62º 21' 00" | N | HESS J1507-622(T) | PWN? | 1? | | | 0.01(≥1TeV)(H) | | 79, 80, 115 |
| 46 | G319.62+0.29 | 15h 03m 38s | -58h 13m 45s | N | HESS J1503-582(T) | | | | | 0.06(≥1TeV)(H) | | |
| 47 | G321.9-0.3 | 15h 20m 40s | -57º 34' 00'' | N | 2FGL J1521.8-5735（G） | S | | | 23'-31' | 0.049(F) | | 17 |
| 48 | G326.3-1.8(MSH 15-56) | 15h 53m 00s | -56º 10' 00'' | N | 2FGL J1552.8-5609(G) | C PWN | | ≥ 1.5(S) | 38' | 0.027(F) | | 17 |
| 49 | G332.4+0.1 (Kes 32, MSH 16-51) | 16h 15m 20s | -50º 42' 00'' | N | 2FGL J1615.0-5051(G) | S | 0.3-3 (S) | 7-11(S) | 15' | 0.066(F) | | 17 |
| 50 | G336.38+0.19 | 16h 32m 09.6s | -47º 49' 12'' | N | HESS J1632-478(T) | PWN | | | | 0.12(≥0.2TeV)(H) | | |
| 51 | G336.7+0.5 | 16h 32m 11s | -47º 19' 00'' | N | 2FGL J1631.7-4720c | S | | | 10'-14' | 0.034(F) | | 17 |
| 52 | G343.0-6.0(RCW 114) | 17h 25m 00s | -46º 30' 00'' | N | 2FGL J1727.3-4611(G) | S | | | 250' | 0.005(F) | | 17 |
| 53 | G343.10-2.68? | 17h 09m 42.2s | -44º 28' 57'' | N | PSR B1706-44(T?) | | | 2.5 | | ≤0.02(≥0.5TeV)(H,C) | | |
| 54 | G350.1-0.3 | 17h 17m 40s | -37º 24' 00'' | N | 2FGL J1718.1-3725(G) | S? | 0.9(S) | 4.5(S) | 4' | 0.018(F) | | 17, 116 |
| 55 | G353.44-0.13 | 17h 29m 35s | -34º 32' 22'' | N | HESS J1729-345(T) | | | | | ?(H) | | 100 |
| 56 | G355.4+0.7 | 17h 31m 20s | -32º 26' 00'' | N | 2FGL J1731.6-3234c (G) | S | | | 25' | 0.027(F) | | 17 |
| 57 | G356.3-0.3 | 17h 37m 56s | -32º 16' 00'' | N | 2FGL J1737.2-3213(G) | S | | | 7'-11' | 0.031(F) | | 17 |
| 58 | G357.7-0.1( | 17h 40m 29s | -30º 58' 00'' | N | 2FGL J1740.4-3054 | S? | | 11.8(S) | 3'- 8' | 0.052(F) | Y | 17 |
| | The Tornado, MSH 17-39) | | | | c(G) | | | | | | | |
| 59 | G358.4+0.19 | 17h 41m 00s | -30º 12' 00" | N | HESS J1741-302(T) | | | | | 0.01(≥1TeV)(H) | | |
| 60 | G358.5-0.9 | 17h 46m 10s | -30º 40' 00" | N | 2FGL J1745.5-3028c(G) | S | | | 17' | 0.023(F) | | 17 |
| 61 | G359.1+0.9 | 17h 39m 36s | -29º 11' 00" | N | 2FGL J1738.9-2908(G) | S | | | 11'-12' | 0.037(F) | | 17 |

Table 2 has the same caption as Table 1.

3. The Prospect of CR originating from SNRs

Next generation very-high-energy telescope arrays will detect more Galactic TeV gamma-ray sources. E.g. LHAASO (see other papers on LHAASO in the special supplement).

It will be expected to identify more faint SNR / TeV γ -ray source associations by employing new generational multi-waveband survey data (e.g., radio survey from the new Australian Square Kilometer Array (ASKAP) telescope) so that we may test these ideas to solve the CR origin puzzle. We have been involved in next generation radio telescope survey programs, e.g. The Galactic Australian Square Kilometer Array Pathfinder survey (GASKAP, also see link for the detail: **http://www.astro.keele.ac.uk/~jacco/research/GASKAP.pdf**). GASKAP is a survey of emission and absorption at 21 cm (the HI line) and of maser emission, diffuse emission, etc. It is an order of magnitude deeper than any survey at comparable resolution (e.g. International Galactic Plane Survey (composed of VLA Galactic Plane Survey (VGPS), CGPS and SGPS, one of the excellent surveys currently available. See Fig. 2). It covers an order of magnitude more area at low latitudes than any survey at comparable sensitivity (e.g. The Galactic Arecibo L-Band Feed Array HI Survey (GALFA), see Fig. 3).

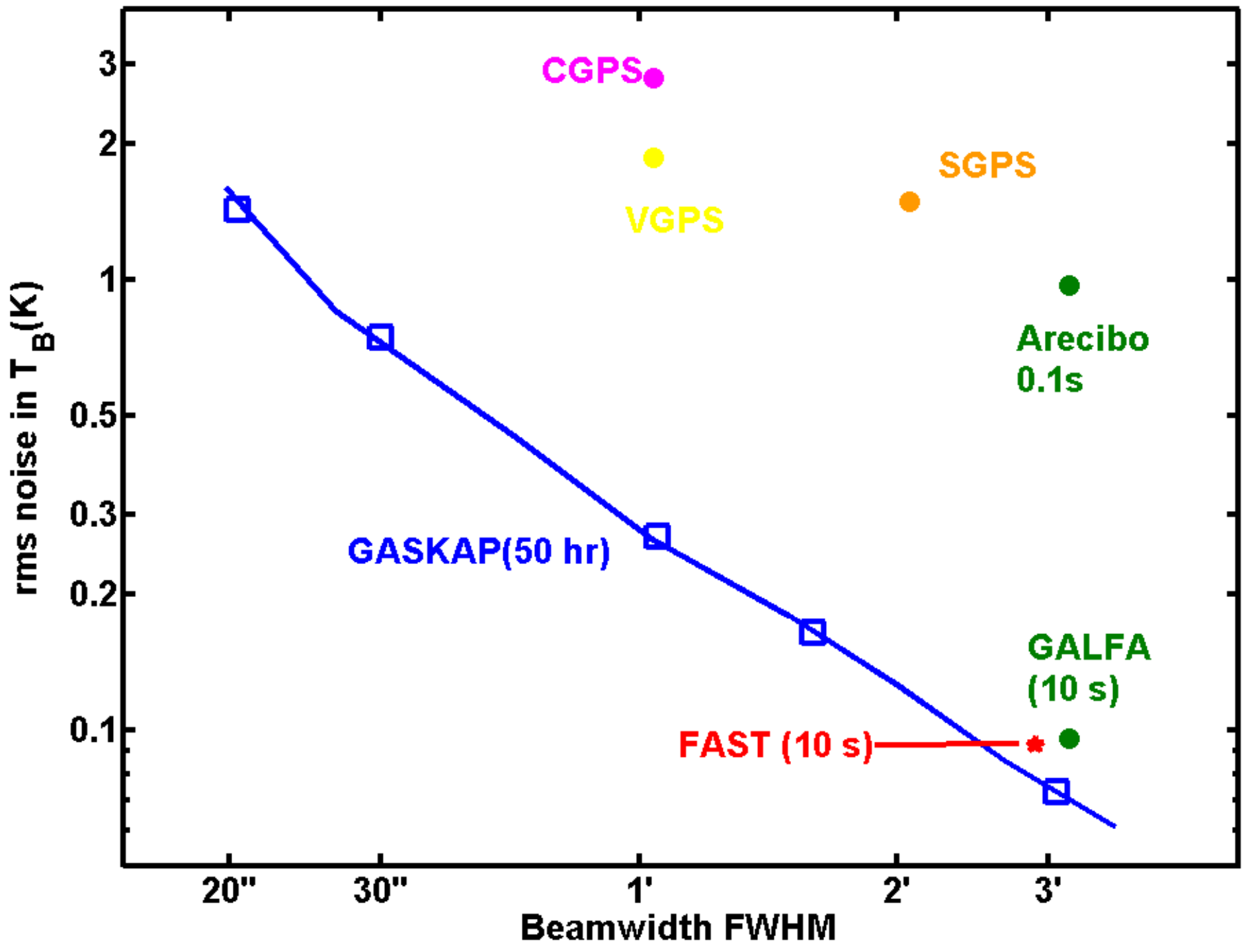

Fig. 2. (Color online)The GASKAP brightness sensitivity vs. resolution (beamwidth), for the medium integration time of 50 hours, with spectra smoothed to 1 km/s. The other two survey speeds give sensitivities a factor of two higher or lower. H I emission mapping at low latitudes will make use of resolution from 20”to 1”, depending on the brightness and angular scales of the emission in each field. FAST is the Chinese next general 500 m single-dish radio telescope. We obtain this Figure by editing Figure 5 in our GASKAP proposal.

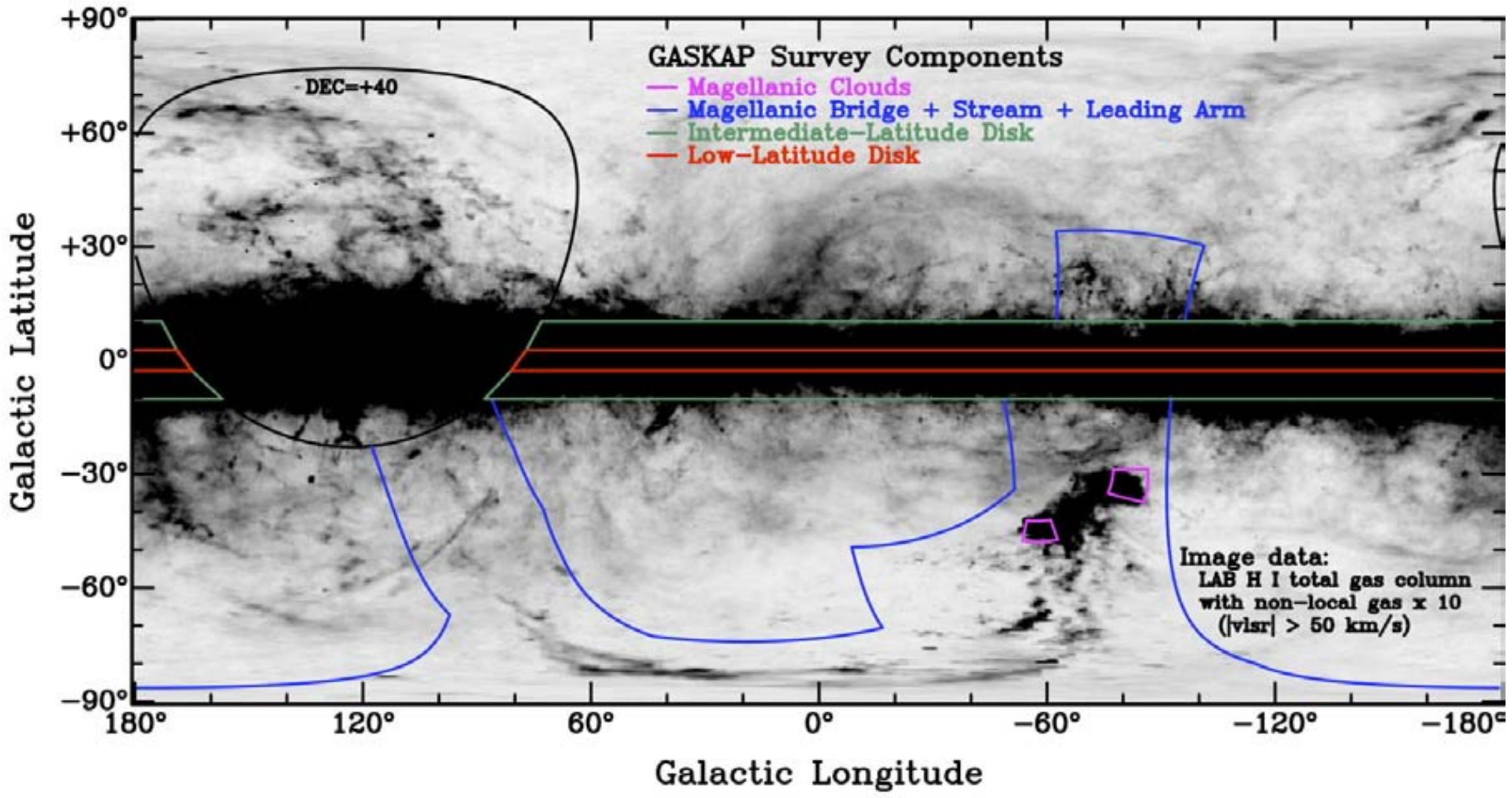

Fig. 3. (Color online)The GASKAP survey areas in Galactic coordinates, with H I column densities (grey scale) in the background. The region north of DEC = +40 Degree must be filled in from the Northern Hemisphere. Fig. 3 is from Fig. 4 of our GASKAP proposal.

ASKAP will start to run in 2013. We expect that we should detect many faint Galactic SNRs by employing ASKAP. We had previously developed distance-determination methods by giving results from HI and CO line spectra of some SNRs[117,118], sowe can use GASKAP data to determine distances to the faint SNRs associated with TeV sources in order to obtain their basic parameters such as the luminosity and age . By combining the deeper very-high-energy-telescopes surveys, we believe that our research may advance understanding of the CRs' origin.

We acknowledge supports from the NSFC program (011241001, 11261140641). This publication was made possible through the support of a grant from the John Templeton Foundation and National Astronomical Observatories (NAO) of the CAS. The opinions expressed in this publication are those of the authors and do not necessarily reflect the views of the John Templeton Foundation of NAOCAS. The funds from John Templeton Foundation were grant from the University of Chicago which also manages the program in conjunction with NAO CAS.